\documentclass[a4paper]{article}

\usepackage{INTERSPEECH2022}
\usepackage{multirow}
\usepackage[super]{nth}
\usepackage{caption}
\usepackage{subcaption}
\usepackage{url}
\usepackage{cite}
\title{Expressive, Variable, and Controllable Duration Modelling in TTS}
\name{
    Ammar Abbas, 
    Thomas Merritt, 
    Alexis Moinet,
    Sri Karlapati, 
    Ewa Muszynska, \\
    Simon Slangen$^\dagger$\thanks{$^\dagger$ The work was done while author was at Amazon.}, 
    Elia Gatti,
    Thomas Drugman
}

\address{
  Alexa AI, Amazon
}

\email{syeabbs@amazon.co.uk}

\begin{document}

\maketitle
\begin{abstract}
\vspace{-0.05cm}

Duration modelling has become an important research problem once more with the rise of non-attention neural text-to-speech systems. 
The current approaches largely fall back to relying on previous statistical parametric speech synthesis technology for duration prediction, which poorly models the expressiveness and variability in speech.
In this paper, we propose two alternate approaches to improve duration modelling. First, we propose a duration model conditioned on phrasing that improves the predicted durations and provides better modelling of pauses. We show that the duration model conditioned on phrasing improves the naturalness of speech over our baseline duration model.
Second, we also propose a multi-speaker duration model called Cauliflow, that uses normalising flows to predict durations that better match the complex target duration distribution. Cauliflow performs on par with our other proposed duration model in terms of naturalness, whilst providing variable durations for the same prompt and variable levels of expressiveness. 
Lastly, we propose to condition Cauliflow on parameters that provide an intuitive control of the pacing and pausing in the synthesised speech in a novel way.

\end{abstract}

\noindent\textbf{Index Terms}: neural text-to-speech, normalising flows, expressive TTS, duration modelling

\vspace{-0.05cm}
\section{Introduction}

Text-to-speech (TTS) systems have seen many paradigm shifts in recent years. Neural TTS systems achieved state-of-the-art results in speech synthesis with different architectures~\cite{wang2017tacotron, shen2018natural, ping2018deep, oord2016wavenet, kalchbrenner2018efficient}. However, they lack robustness to out-of-distribution text, thus causing issues such as mumbling, skipping, or repetition in speech. Therefore, various new methods have been proposed that are more robust to unseen text utterances~\cite{yu2019durian, ren2019fastspeech}. They employ duration models to find alignment between input phonemes and target acoustic features. Many recent methods~\cite{zeng2020aligntts, ren2020fastspeech, elias2021parallel} have been based on this architecture, and have gradually improved the acoustic models. At the same time, the duration models have largely not been updated since the statistical parametric speech synthesis approaches. 
These duration models are deterministic, and generally trained via distance-based losses like L1/L2 loss without any additional conditioning. This results in close to mean duration prediction in speech with limited expressiveness. These uni-modal predicted durations also lead to insufficient pauses in longer texts, adversely affecting the listener's understanding of the text~\cite{rochester1973significance}. Klimkov~\emph{et~al.}~\cite{klimkov2017phrase} and Liu~\emph{et~al.}~\cite{liu2020modeling} share a phrasing model that is introduced to address this shortcoming, however, they are presented for an attention based TTS system.

Variability in the synthesised speech is another important factor in order to improve the perceived naturalness of the TTS systems for long and repetitive texts such as question-answering prompts.
This is achievable using some of the recent generative modelling techniques such as normalising flows also known as flows. Flows~\cite{kingma2018glow} provide a tractable solution for maximising exact log-likelihood which allows predicting distributions that are true to the target distribution. Furthermore, they allow variations in the output by controlling the prior. Flows have recently attracted attention in the TTS domain for modelling diverse acoustic features~\cite{kim2020glow, miao2020flow, valle2020flowtron}. However, the duration models used are still deterministic, and trained with a distance-based loss, thus suffering from the aforementioned limitations. Kim~\emph{et~al.}~\cite{kim2021conditional} describe a similar approach to ours by using a stochastic duration model, however, they obtain the target durations for their duration model via Monotonic Alignment Search (MAS). MAS can suffer from attention instabilities when extracting the target durations during training~\cite{kim2020glow}. Additionally, their duration model is conditioned only on phoneme embeddings, and has a significantly different architecture.

In this paper, we present two alternate duration modelling approaches: (a) duration model conditioned on phrasing; (b) Cauliflow. Our main contributions are: (i) We show that duration model conditioned on phrasing is statistically significantly preferred to the baseline duration model; (ii) We explore the performance of the duration model conditioned on phrasing, and show that it improves the predicted durations, pause rate, speech rate, and $\mathcal{F}_0.25$ score over the baseline duration model; (iii) We propose a flow-based multi-speaker stochastic duration model called Cauliflow (Conditional Alignment Using normaLIsing FLOWs) that is conditioned on BERT word embeddings to predict semantic and syntax informed durations. Cauliflow performs on par with duration model conditioned with phrasing in terms of naturalness; and finally (iv) We show that Cauliflow predicts more expressive durations, and can produce variable durations for the same text. 
Additionally, we show that the proposed novel parameters in Cauliflow control the pace and pausing in the speech in an intuitive way.

\begin{figure*}
     \centering
     \begin{subfigure}[b]{0.3\linewidth}
         \centering
         \includegraphics[width=\textwidth]{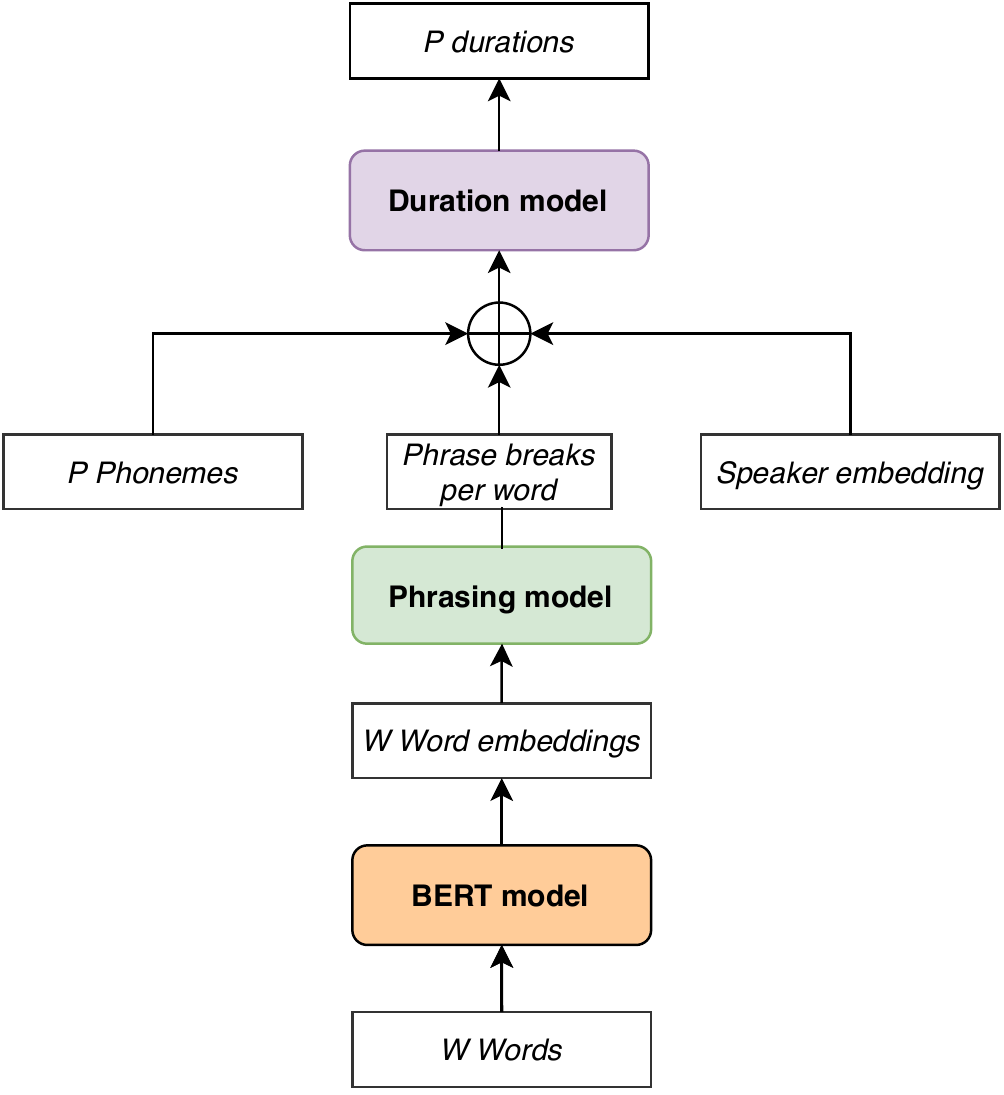}
         \caption{Duration model conditioned on phrasing}
         \label{fig:phrasing_model}
     \end{subfigure}
     \hfill
     \begin{subfigure}[b]{0.35\linewidth}
         \centering
         \raisebox{5mm}{\includegraphics[width=\textwidth]{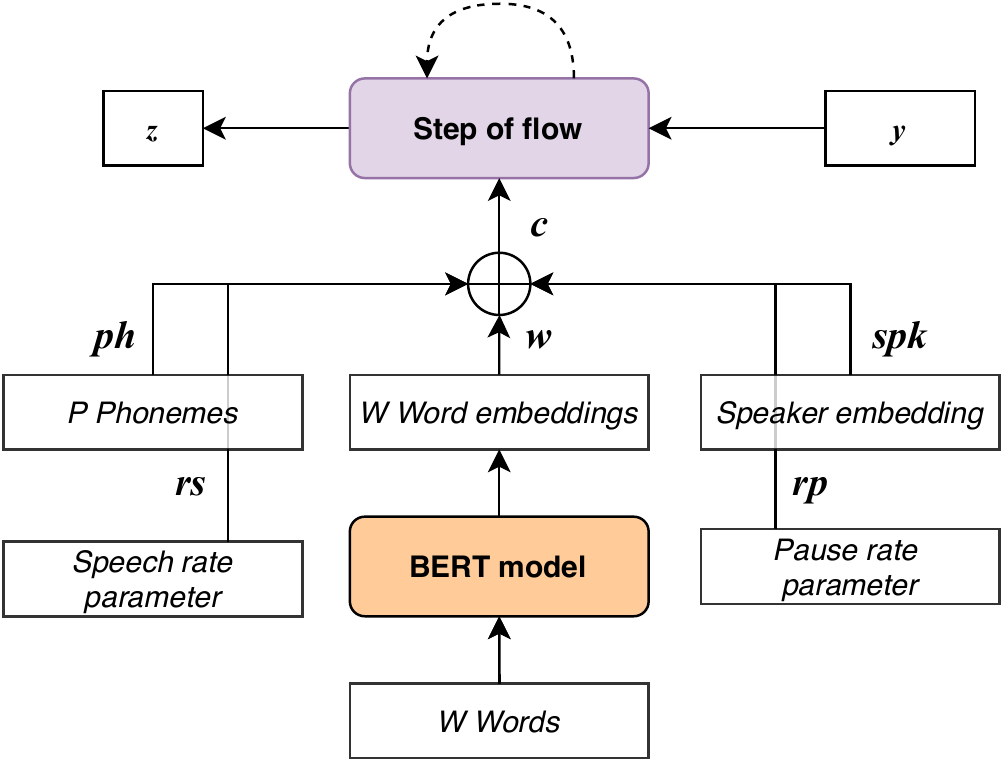}}
         \caption{Cauliflow model}
         \label{fig:cauliflow_architecture}
     \end{subfigure}
     \hfill
     \begin{subfigure}[b]{0.27\linewidth}
         \centering
         \includegraphics[width=\textwidth]{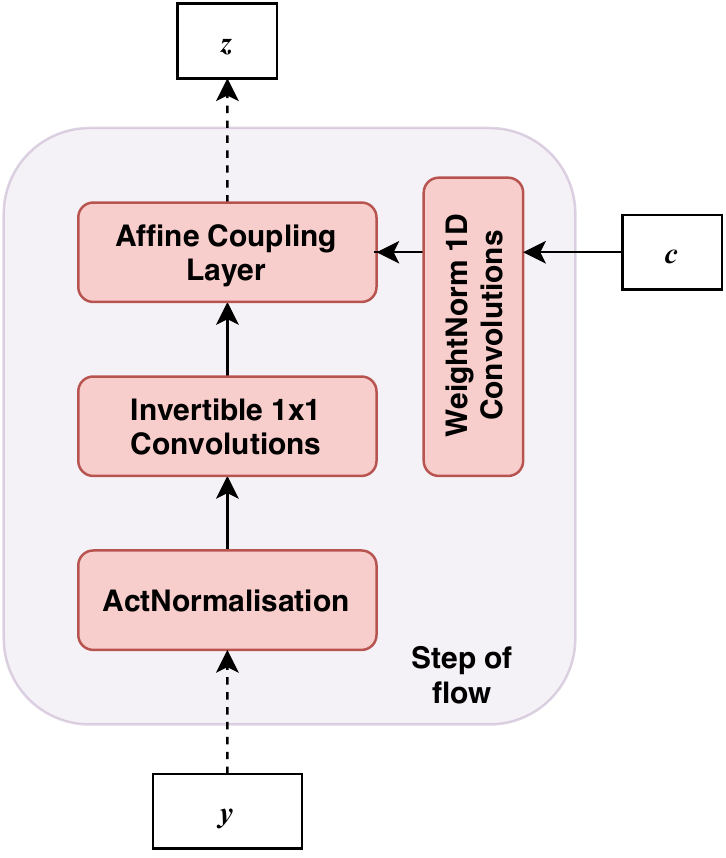}
         \caption{Inside view of a step of flow}
         \label{fig:step_of_flow}
     \end{subfigure}
     \hfill
     \vspace{-0.1cm}
        \caption{Comparison of the architecture of Cauliflow and duration model conditioned on phrasing. (b) and (c) shown at training time.}
        \label{fig:three_architectures}
        \vspace{-0.4cm}
\end{figure*}

\section{Baseline}
\label{sec:baseline}
  
\subsection{Acoustic Model}

We train an acoustic model that is a multi-speaker version of the baseline introduced in \cite{karlapati2021prosodic}.
The acoustic model is trained separately from the duration model. This allows us to compare different duration models given the same acoustic model.
First, we obtain $P$ phoneme embeddings $\vec{ph}=[{ph}_0, {ph}_1, \cdots, {ph}_{P-1}]$ from the encoder given input text, where $P$ is the number of phonemes. The phoneme embeddings $\vec{ph}$ are then upsampled according to the durations $\vec{d}=[d_0, d_1, \cdots, d_{P-1}]$, such that $\sum_{i=0}^{P-1}{d_i} = T$, where $T$ is the total number of mel-spectrogram frames. 
The upsampled phoneme embeddings are passed through the decoder to obtain the mel-spectrogram. We use a universal neural vocoder~\cite{jiao2021universal} to obtain speech signal from the generated mel-spectrogram.

\subsection{Duration model}
\label{subsec:duration}
\vspace{-0.1cm}

We train a duration model that is optimised with the L2 loss to predict normalised durations $\vec{d}$, a common approach in many duration based architectures~\cite{ren2020fastspeech, kim2020glow, miao2020flow}. All duration models in this paper are trained using the same oracle durations, obtained from Montreal Forced Alignment~\cite{mcauliffe2017montreal} for a fair comparison.

\section{Duration model conditioned on phrasing}
\label{subsec:duration_with_phrasing}

The baseline duration model is trained with the L2 loss that has some limitations in practice. The local minimum converges to the mean of the data~\cite{zhao2016loss} which can result in monotonous durations if any additional conditioning is not provided, thus producing flatter or less expressive speech. This particularly affects the prediction of the occurrence and duration of pauses.
Indeed, in our data, most pauses occur at punctuation marks, but more than $20\%$ can be found between words not separated by a punctuation mark, i.e. a word boundary. However, only around $2.5\%$ of word boundaries correspond to a pause. 
Thus due to the averaging nature of the L2 loss, our baseline model predicts that the duration at any word boundary approaches 0 (i.e. no pause), unless there is a punctuation mark. 

To better model the distribution of pauses in the data, we train a duration model that is conditioned on output from a phrasing model. The phrasing model is a classification model trained with Cross Entropy loss against target binary phrasing labels that are obtained from forced alignment.
The phrasing model architecture can be seen in Figure~\ref{fig:phrasing_model}. The input text is passed through a fine-tuned BERT-base model~\cite{kenton2019bert}.
The BERT model predicts sub-word level embeddings from the input text, which are then averaged to compute word-level BERT embeddings passed to the phrasing model.

The output of the phrasing model is the probability of a pause after each word that is converted into a binary value, i.e., occurrence of a pause, before being passed to the duration model. 
We choose a threshold that optimises the $\mathcal{F}_{0.25}$ score~\cite{klimkov2017phrase} for predicted pauses in order to obtain higher precision for a better listening experience. During training, the duration model is conditioned on oracle phrasing labels. During inference, it is conditioned on the output of the phrasing model.

\section{Cauliflow}
\label{sec:cauliflow}

Cauliflow is a normalising flow based duration model that is conditioned on phoneme, BERT and speaker embeddings. Additional conditioning on speech and pause rate allows us to further control the predicted duration distribution.

Normalising flows allow optimising for the exact log-likelihood of the target distribution. 
Let $\vec{z}$ be the prior Gaussian distribution with the probability density function $P(\vec{z})$, and let $P(\vec{y}|\vec{c})$ be the target duration $\vec{y}$ distribution given conditional features $\vec{c}$. Normalising flows are composed of series of invertible functions $\bm{f}(\vec{z}|\vec{c}) = f_0 \circ f_1 \circ f_2 \cdots f_k$ that transform the Gaussian distribution $\vec{z}$ into the target distribution $\vec{y}$:

\vspace{-0.1cm}
\begin{equation}
\label{eq:composition_transform}
    \vec{y} = \bm{f}(\vec{z}|\vec{c}) = f_0 \circ f_1 \circ f_2 \cdots f_k(\vec{z}|\vec{c}),
\end{equation}

then by a change of variables, we can compute the log-likelihood of the target distribution $\vec{y}$ as:

\vspace{-0.1cm}
\begin{equation}
    \log P_Y(\vec{y}|\vec{c}) = \log P_Z(\vec{z}) + \sum_{i=1}^{K} \log|\det( \bm{J}(f^{-1}_i(\vec{x}|\vec{c})))|,
\end{equation}
\vspace{-0.3cm}

where $\log|(\det(\bm{J}))|$ represents the log-value of the determinant of the Jacobian matrix $\bm{J}$.

\begin{table*}[tb]
\centering
\begin{tabular}{|c|l|rr|rrr|rrr|r|r|}
\hline
\multirow{2}{*}{\textbf{Speaker}} &
  \multicolumn{1}{c|}{\multirow{2}{*}{\textbf{System}}} &
  \multicolumn{2}{c|}{\textbf{JSD}} &
  \multicolumn{3}{c|}{\textbf{Punctuations}} &
  \multicolumn{3}{c|}{\textbf{Word boundaries}} &
  \multicolumn{1}{c|}{\multirow{2}{*}{\textbf{Pause rate}}} &
  \multicolumn{1}{c|}{\multirow{2}{*}{\textbf{Speech rate}}} \\ \cline{3-10}
 &
  \multicolumn{1}{c|}{} &
  \multicolumn{1}{c|}{\textbf{Pause}} &
  \multicolumn{1}{c|}{\textbf{Non-pause}} &
  \multicolumn{1}{c|}{$\mathcal{P}$} &
  \multicolumn{1}{c|}{$\mathcal{R}$} &
  \multicolumn{1}{c|}{$\mathcal{F}_{0.25}$} &
  \multicolumn{1}{c|}{$\mathcal{P}$} &
  \multicolumn{1}{c|}{$\mathcal{R}$} &
  \multicolumn{1}{c|}{$\mathcal{F}_{0.25}$} &
  \multicolumn{1}{c|}{} &
  \multicolumn{1}{c|}{} \\ \hline
\multirow{4}{*}{A} & \textbf{Target}    &   -  &    - &     - &     - &     - &     - &     - &     - & 5.86  & 3.07 \\ \cline{2-2}
                   & \textbf{Dur}       & 0.36 & 0.09 & 92.27 & 92.78 & 92.30 & 72.28 & 12.71 & 56.65 & 6.98  & 3.32 \\ \cline{2-2}
                   & \textbf{Dur+P}     & 0.25 & 0.07 & 91.07 & 98.66 & 91.48 & 65.63 & 64.32 & 65.55 & 5.66  & 3.10 \\ \cline{2-2}
                   & \textbf{Cauliflow} & 0.19 & 0.03 & 95.01 & 83.66 & 94.25 & 66.58 & 46.82 & 64.97 & 6.30  & 3.11 \\ \hline
\multirow{4}{*}{B} & \textbf{Target}    &     - &    - &     - &     - &    -  &     - &     - &     - & 7.65  & 3.59 \\ \cline{2-2}
                   & \textbf{Dur}       & 0.56 & 0.09 & 82.29 & 48.19 & 79.00 & 0.0   & 0.0   & 0.0   & 10.53 & 3.94 \\ \cline{2-2}
                   & \textbf{Dur+P}     & 0.26 & 0.07 & 76.93 & 95.52 & 77.82 & 42.80 & 48.13 & 43.08 & 6.85  & 3.48 \\ \cline{2-2}
                   & \textbf{Cauliflow} & 0.15 & 0.04 & 82.76 & 70.59 & 81.93 & 33.21 & 37.76 & 33.45 & 7.81  & 3.46 \\ \hline
\end{tabular}
\caption{Comparison of objective metrics between different durations models. Pause and speech rate are better when closer to Target.}
\label{table:objective_metrics_all}
\vspace{-0.6cm}
\end{table*}

Cauliflow is specifically based on the architecture of Flow-TTS~\cite{miao2020flow}. An overview of Cauliflow model can be seen in Figure~\ref{fig:cauliflow_architecture}. 
The conditions $\vec{c}$ are added in the affine coupling layers~\cite{dinh2016density} in the flow steps as shown in Figure~\ref{fig:step_of_flow}, according to equation~\ref{eq:composition_transform}. We now detail the steps required to compute different parameters in $\vec{c}$.

\vspace{-0.15cm}
\subsection{Phoneme embeddings}
\vspace{-0.1cm}

The phoneme embeddings $\vec{ph}$ contain information about the content to be synthesised. They are obtained by passing the phonemes $\vec{p}$ through an encoder with an architecture similar to the encoder used in Tacotron 2~\cite{shen2018natural}.

\vspace{-0.15cm}
\subsection{BERT embeddings}
\label{subsubsec:bert_embeddings}
\vspace{-0.1cm}

The semantic and syntactical information in a sentence plays an important role in the prosody of the synthesised speech~\cite{bennett2019syntax, nygaard2009semantics}. BERT embeddings have been shown to contain information about the semantic and syntactical structure of a sentence~\cite{rogers2020primer}. To leverage that information to guide duration prediction, we condition Cauliflow on embeddings obtained from BERT. First, we obtain word-level BERT embeddings using the process described in Section~\ref{subsec:duration_with_phrasing}. The word-level BERT embeddings are then upsampled according to the number of phonemes in each word to compute phoneme-level BERT embeddings $\vec{w}$.

\vspace{-0.15cm}
\subsection{Speaker embeddings}
\vspace{-0.1cm}

The speaker embeddings $\vec{spk}$ are obtained from a speaker verification system that is trained using a network based on Generalised End-to-End loss~\cite{wan2018generalized}. A speaker embedding is a 192-dimensional vector computed for each utterance during training. An utterance can be composed of one or more sentences. During inference, we use  a  mean speaker embedding that is computed given all the training utterances for a particular speaker.

\vspace{-0.15cm}
\subsection{Speech and pause rate}
\vspace{-0.1cm}

A parameter $\vec{rs}$ is introduced to control the pace of the speech. Let $\mu_{rs}$ be the average number of words per second in the training data. Then the model is conditioned on $\vec{rs}$, calculated for each utterance as:

\vspace{-0.1cm}
\begin{equation}
    \vec{rs} = \frac{W}{D} - \mu_{rs},
\end{equation}

where $W$ is the number of words, and $D$ is the duration of the utterance in seconds. 
The model learns the relation between $\vec{rs}$  and the speech rate to be controlled in the synthesised speech, i.e., the scalar value 0 represents the average number of words per second seen in the data, while positive or negative values of $\vec{rs}$ represent a faster or slower speech respectively. A similar approach is followed for the parameter $\vec{rp}$ controlling the pause rate which is calculated as:

\vspace{-0.1cm}
\begin{equation}
    \vec{rp} = \frac{W}{S} - \mu_{rp},
\end{equation}

where $S$ is the number of pauses in the utterance, and $\mu_{rp}$ is the average number of words per pause in an utterance in the training data.

Given these different parameters, Equation~\ref{eq:composition_transform} for normalising flows can be expanded to:

\vspace{-0.1cm}
\begin{equation}
    \vec{y} = f(\vec{z}|\vec{c}) =  f(\vec{z}|\vec{ph}, \vec{w}, \vec{spk}, \vec{rp}, \vec{rs})
\end{equation}

The target durations $\vec{y}$ are not normalised in Cauliflow. We experimented with different normalisation schemes such as log-normalisation and z-score, but found that non-normalised target durations give the best results.
This could be due to possible outliers which are discussed in Section~\ref{subsubsec:control_temperature}.

\vspace{-0.05cm}
\section{Experimental Validation}
\vspace{-0.05cm}

We evaluate three different duration models: a) the baseline from Section~\ref{subsec:duration}, referred as ``Dur''; b) duration model conditioned on phrasing from Section~\ref{subsec:duration_with_phrasing}, referred as ``Dur+P''; c) Cauliflow from Section~\ref{sec:cauliflow}. The ground truth durations are referred as ``Target''.

\vspace{-0.15cm}
\subsection{Data}
\vspace{-0.1cm}

Experiments are conducted on an internal English voice dataset consisting of 1 male and 3 female speakers. The models are trained on all 4 voices comprising 50 hours of total training data, while the test set consists of around 3 hours of data for each speaker. We report our results on a subset of two speakers chosen randomly, referred from hereon as speaker A and speaker B. The sampling rate of the recorded audio is 24 kHz. We extract 80 band mel-spectrograms from the audio with a frame shift of 12.5ms.

\vspace{-0.15cm}
\subsection{Comparing Dur+P against Dur}
\label{subsec:phrasing_model_eval}
\vspace{-0.1cm}

\begin{figure}[tb]
  \centering
  \includegraphics[width=0.8\linewidth]{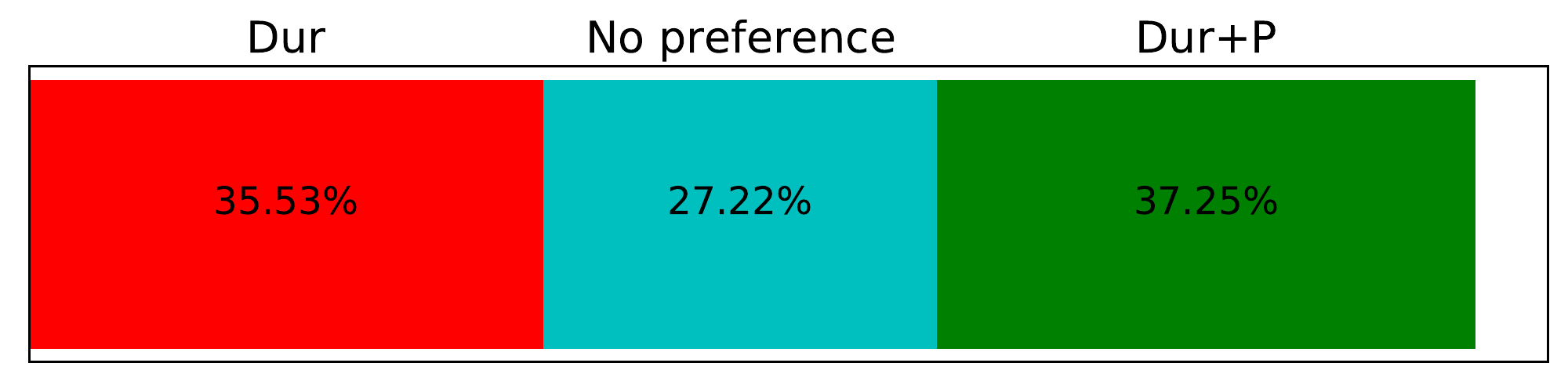}
  \vspace{-0.2cm}
  \caption{Preference test result shows that \text{Dur+P} is statistically signficantly better than \text{Dur} on speaker A $(\text{p-val}<0.05)$.}
  \label{fig:pref_durp}
  \vspace{-0.1cm}
\end{figure}

We conducted a preference test between Dur and Dur+P. The evaluation consisted of 50 utterances randomly selected from speaker A test set, and were rated by 120 native English listeners. The results are shown in Figure~\ref{fig:pref_durp}. We use a binomial significance test to measure the statistical significance, and find that Dur+P is statistically significantly better than Dur (p-val\textless{}0.05).

To gain more insights, we also compare the durations predicted by both models on the whole test set as shown in Table~\ref{table:objective_metrics_all}, using different objective metrics such as speech and pause rate, precision $\mathcal{P}$, recall $\mathcal{R}$, and $\mathcal{F}_{0.25}$ score~\cite{klimkov2017phrase} over pauses, and Jensen-Shannon divergence (JSD). We choose JSD as a metric because it measures the closeness between two distributions in a symmetric way. The comparison shows that Dur+P has a speech and pause rate that is closer to the speaker's statistics. 
The Dur model does not predict a pause on any word boundary for speaker B, while Dur+P significantly improves the ${F}_{0.25}$ for both speakers. We also note that Dur+P has lower JSD on predicted duration distribution for both speakers on pause and non-pause tokens. We believe that the model is also performing better on non-pause tokens because information about pause occurrence can affect the duration of adjacent phonemes.

\begin{table}[tb]
\begin{tabular}{|l|lll|}
\hline
                   & \textbf{Dur+P} & \textbf{Cauliflow} & \textbf{CopySynth} \\ \hline
\textbf{Speaker A} & 73.58±1.10     & 73.53±1.10         & 76.03±1.02         \\
\textbf{Speaker B} & 73.62±1.13     & 73.22±1.14         & 74.21±1.12         \\ \hline
\end{tabular}
\caption{Comparison of MUSHRA scores between different systems along with their 95\% confidence intervals.}
\label{table:mushra_scores}
\vspace{-0.8cm}
\end{table}

\vspace{-0.15cm}
\subsection{Comparing Cauliflow against Dur+P}
\vspace{-0.1cm}

In this section, we evaluate both of our proposed duration models, Cauliflow and Dur+P. We run a MUSHRA~\cite{series2014method} evaluation consisting of Cauliflow, Dur+P, and copy-synthesis of the original recordings (CopySynth) to quantify improvements in naturalness of the systems. 
Copy-synthesis is used to focus only on improvements from the duration model and its effect on the acoustic model.
The evaluations are run for speaker A and B consisting of 50 utterances each rated by a total of 240 native English speakers. The results of the evaluations can be seen in Table~\ref{table:mushra_scores}. We run a pairwise two-sided Wilcoxon signed-rank test, corrected for multiple comparisons to measure statistical significance between the systems. 
We found that Cauliflow and Dur+P are on par in terms of naturalness on both speakers $(\text{p-val}>0.05)$. Both the systems obtain MUSHRA scores that are very close to the MUSHRA score for CopySynth on speaker B, with no statistically significant difference between Dur+P and CopySynth.

\begin{figure}[t]
  \centering
  \includegraphics[width=0.8\linewidth]{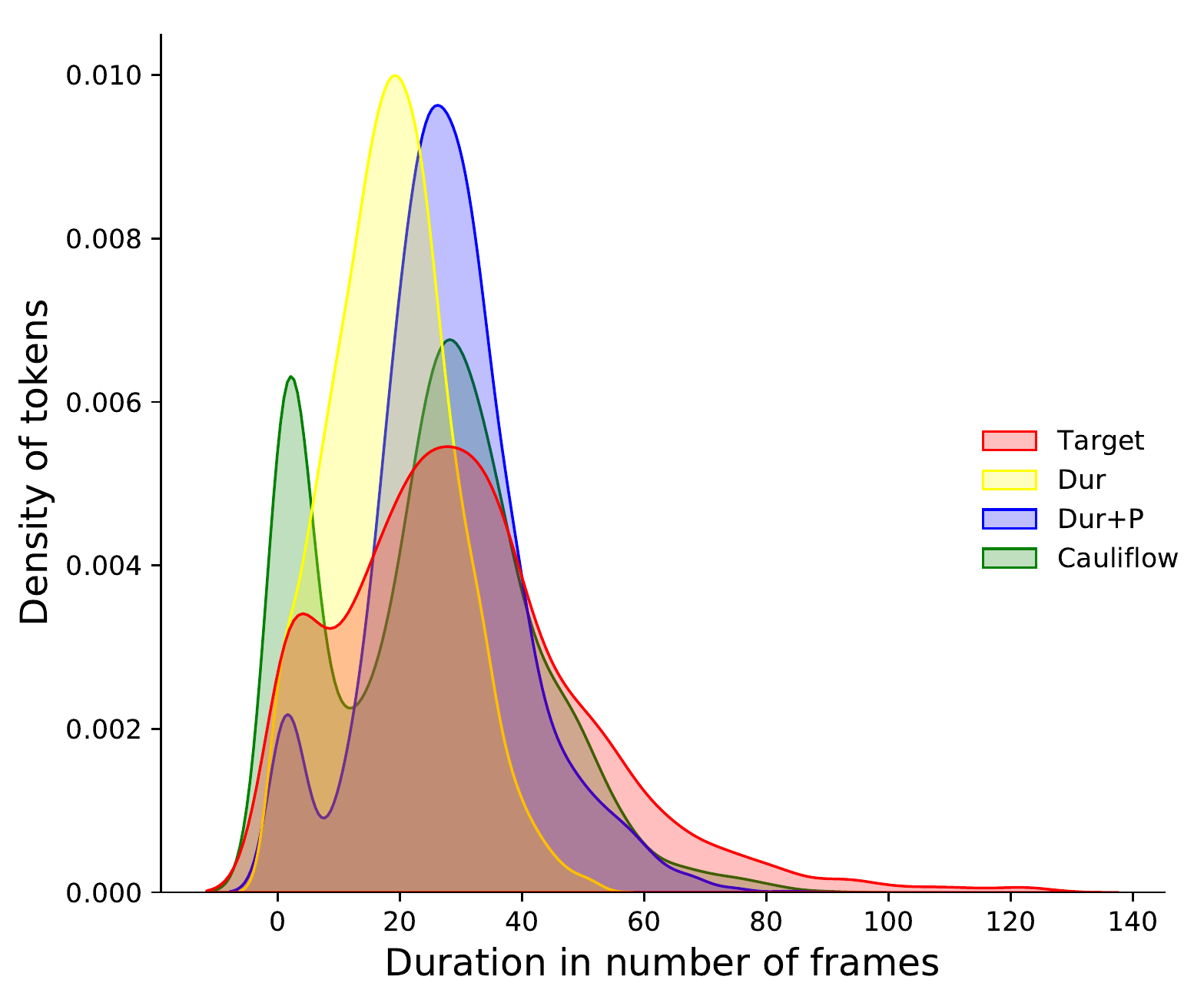}
  \vspace{-0.2cm}
  \caption{Comparison of predicted durations on punctuation marks by different duration models.}
  \label{fig:on_puncs}
  \vspace{-0.1cm}
\end{figure}

To understand more about how the predicted durations differ between both the models, we show their objective metrics in Table~\ref{table:objective_metrics_all}. 
We note that Cauliflow generally has a better precision, while Dur+P has a better recall on predicted pauses, thus providing use cases for different situations.
The predicted durations from Cauliflow have lower JSD with the target durations compared to Dur+P on both pauses and non-pause phonemes for both speakers. 
This can also be seen by the duration distribution of these methods over the punctuation tokens in Figure~\ref{fig:on_puncs}. The Dur+P model has a higher mode whereas Cauliflow has a better bi-modal distribution of durations that denotes pauses vs non-pauses.
The tails of the Cauliflow distribution also show better distribution coverage but can sometimes result in outlier predictions. 
This shows that Cauliflow produces a more natural duration distribution but it can suffer from outlier predictions. We further explore the latter in detail in the next section.

\begin{table}[]
\begin{tabular}{|l|rr|r|}
\hline
\multicolumn{1}{|c|}{\multirow{2}{*}{\textbf{System}}} &
  \multicolumn{2}{c|}{\textbf{JSD}} &
  \multicolumn{1}{c|}{\multirow{2}{*}{\textbf{\begin{tabular}[c]{@{}c@{}}\nth{99} Percentile \\ L1 Error \\ (frames)\end{tabular}}}} \\ \cline{2-3}
\multicolumn{1}{|c|}{} &
  \multicolumn{1}{c}{\textbf{Pauses}} &
  \multicolumn{1}{c|}{\textbf{\begin{tabular}[c]{@{}c@{}}Non-\\ pauses\end{tabular}}} &
  \multicolumn{1}{c|}{} \\ \hline
Cauliflow ($\mathcal{T}=0.3$) & 0.21 & 0.05 & 28 \\
Cauliflow ($\mathcal{T}=0.5$) & 0.20  & 0.04 & 29 \\
Cauliflow ($\mathcal{T}=0.7$) & 0.19 & 0.03 & 31 \\
Cauliflow ($\mathcal{T}=1.0$) & 0.19 & 0.03 & 34 \\ \hline
\end{tabular}
\caption{Comparison of duration metrics for various $\mathcal{T}$.}
\label{table:percentile_table}
\vspace{-0.7cm}
\end{table}

\vspace{-0.15cm}
\subsection{Controlling duration variability}
\label{subsubsec:control_temperature}
\vspace{-0.1cm}

We evaluate the effect of temperature $\mathcal{T}$ in the prior Gaussian distribution $\mathcal{N}(0, \sigma.\mathcal{T})$ on the quality of predicted durations from Cauliflow. We use a prior Gaussian distribution with parameters $\mathcal{N}(0, 1)$ during training. However, we note that with $\mathcal{T}=1$, the model predicts durations that can result in unstable speech synthesis. It is partly attributed to the outliers identified by computing different percentile errors. The \nth{99} percentile of the errors in predicted durations vs target durations is shown in Table~\ref{table:percentile_table}. With a lower value of $\mathcal{T}$, we note that there are fewer outliers, however, the model starts to predict worse duration distributions as shown by the higher JSD in the table. We find that $\mathcal{T}=0.7$ is the right trade-off between a better predicted duration distribution and minimizing the outlier predictions. In the the future, we will investigate if there is a way to minimize outlier predictions while still using $\mathcal{T}=1$.

\vspace{-0.15cm}
\subsection{Controlling speech and pause rate via Cauliflow}
\vspace{-0.1cm}

\begin{figure}[t]
  \centering
  \includegraphics[width=0.8\linewidth]{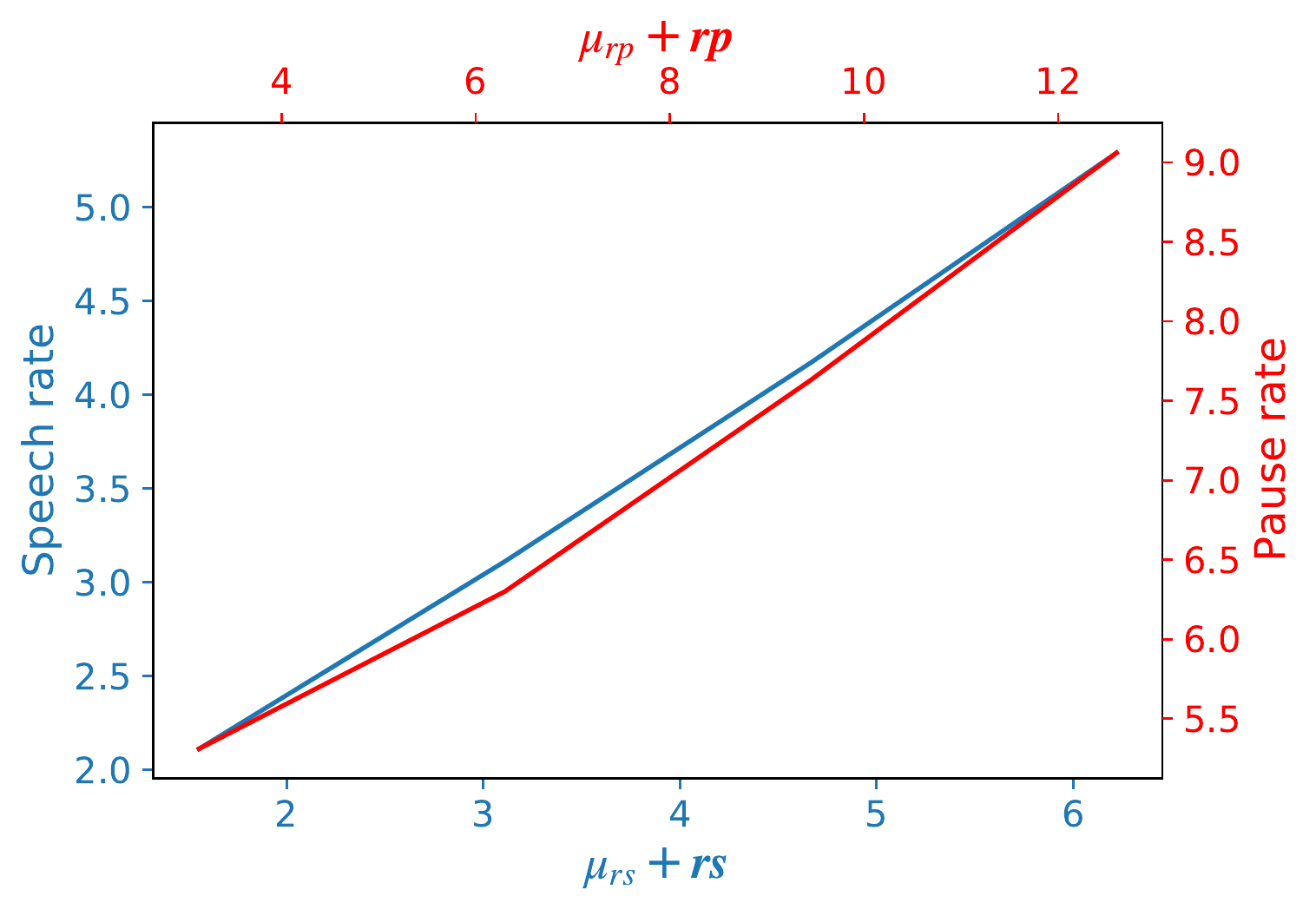}
  \vspace{-0.2cm}
  \caption{Correlation of parameters for speech and pause rate versus their measured impact on speech.}
  \label{fig:speech_and_pause_rate}
  \vspace{-0.1cm}
\end{figure}

We vary the values of $\vec{rp}$ and $\vec{rs}$ to measure their effect on the pace and pausing of the synthesised speech on 3 hours of test set of speaker A. The resultant effect can be seen in Figure~\ref{fig:speech_and_pause_rate}. We find that that there is a direct correlation between control parameters versus the speech and pause rate. Furthermore, the change in pacing and pausing is close to the intended value of the parameters while being a bit conservative, i.e., the decrease or increase in the speech or pause rate is generally less than the intended value. In the case of pause rate, we note that this is generally because the placement of pauses follows the distribution seen in the data, and thus doesn't put pauses at inappropriate locations in the text. An example of varying pause rates for a sentence can be seen in Figure~\ref{table:pauses_same_text}. We vary $\vec{rp}$ decreasing it to as low as around one word per pause, but note that once all semantically appropriate pauses are placed in the text, the model doesn't place any more pauses. Furthermore, new pauses are placed in the decreasing order of their suitability to text ,i.e., pause \textless{}P1\textgreater is placed when using the default $\vec{rp}$, \textless{}P2\textgreater is placed when decreasing it by around 1 word per pause, and so on. Therefore, we believe that these parameters can be safely used in practice as they generally avoid pause insertion that could result in a bad listening experience.

\begin{table}
\centering
\begin{tabular}{|p{0.93\linewidth}|}
\hline
``Are you reading \textless{}P1\textgreater~or just wondering \textless{}P3\textgreater~that whatever happens here \textless{}P4\textgreater~around this time \textless{}P2\textgreater~will stay here forever?'' \\
\hline
\end{tabular}
\caption{Insertion of more pauses in a textual prompt in the order \textless{}P1\textgreater, \textless{}P2\textgreater, \textless{}P3\textgreater, \textless{}P4\textgreater.}
\label{table:pauses_same_text}
\vspace{-0.7cm}
\end{table}

\vspace{-0.05cm}
\section{Conclusion}

In this paper, we proposed a modification to duration models that incorporates a phrasing model, and showed that it improves the predicted durations and produces more natural speech over the baseline. We also proposed a normalising flow based multi-speaker duration model called Cauliflow that is conditioned on BERT embeddings for more appropriate duration prediction. We showed that Cauliflow performs on par with the duration model conditioned on phrasing, while providing variable durations, and providing an intuitive way to control the speech and pause rate of the synthesised speech.

\bibliographystyle{IEEEtran}

\bibliography{refs}

\end{document}